\documentclass[reprint,aps,superscriptaddress,twocolumn,showkeys,showpacs,nofootinbib,longbibliography]{revtex4-2}

\usepackage{graphicx,color}
\usepackage{amsmath,amsfonts,enumerate,amsthm,amssymb,bbm}
\usepackage{float}
\usepackage[colorlinks=true,citecolor=blue,linkcolor=magenta]{hyperref}
\usepackage{multirow}
\usepackage{bbold}
\usepackage{braket}
\usepackage{soul}
\usepackage{mathtools}

\usepackage[version=4]{mhchem} 
\usepackage[caption = false]{subfig}

\usepackage{scrextend}
\usepackage{xcolor}
\usepackage{wrapfig}
\usepackage{cleveref}

\usepackage[switch,columnwise]{lineno}

\long\def\ca#1\cb{}

\newcommand{\abs}[2][]{#1| #2 #1|}

\newcommand{\Tr}{\mathrm{Tr}}

\renewcommand{\geq}{\geqslant}
\renewcommand{\leq}{\leqslant}

\newcommand*{\id}{\openone}

\newcommand{\beginsupplement}{%
        \setcounter{table}{0}
        \renewcommand{\thetable}{S\arabic{table}}%
        \setcounter{figure}{0}
        \renewcommand{\thefigure}{S\arabic{figure}}%
     }
{}
{}

\def\m{^{(m)}}

\begin{document}
\title{Gibbs state sampling via cluster expansions
}
\author{Norhan M. Eassa}
\affiliation{Department of Physics and Astronomy, Purdue University, West Lafayette, IN 47906, USA}
\affiliation{Quantum Science Center, Oak Ridge National Laboratory, Oak Ridge, TN 37831, USA}
\author{Mahmoud M. Moustafa}
\affiliation{Department of Physics and Astronomy, Purdue University, West Lafayette, IN 47906, USA}
\author{Arnab Banerjee}
\email[]{arnabb@purdue.edu}
\affiliation{Department of Physics and Astronomy, Purdue University, West Lafayette, IN 47906, USA}
\affiliation{Quantum Science Center, Oak Ridge National Laboratory, Oak Ridge, TN 37831, USA}
\author{Jeffrey Cohn}
\email[]{jeffrey.cohn@ibm.com}
\affiliation{Quantum Science Center, Oak Ridge National Laboratory, Oak Ridge, TN 37831, USA}
\affiliation{IBM Quantum, IBM Research – Almaden, San Jose, CA 95120, USA}

\begin{abstract} 
    Gibbs states (i.e., thermal states) can be used for several applications such as quantum simulation, quantum machine learning, quantum optimization, and the study of open quantum systems. Moreover, semi-definite programming, combinatorial optimization problems, and training quantum Boltzmann machines can all be addressed by sampling from well-prepared Gibbs states. With that, however, comes the fact that preparing and sampling from Gibbs states on a quantum computer are notoriously difficult tasks. Such tasks can require large overhead in resources and/or calibration even in the simplest of cases, as well as the fact that the implementation might be limited to only a specific set of systems. We propose a method based on sampling from a quasi-distribution consisting of tensor products of mixed states on local clusters, i.e., expanding the full Gibbs state into a sum of products of local “Gibbs-cumulant” type states easier to implement and sample from on quantum hardware. We begin with presenting results for 4-spin linear chains with XY spin interactions, for which we obtain the $ZZ$ dynamical spin-spin correlation functions. We also present the results of measuring the specific heat of the 8-spin chain Gibbs state $\rho_8$.
\end{abstract}
\maketitle
\section{Introduction} 
Gibbs states are mixed quantum states that describe quantum systems at thermodynamic equilibrium with their environment at a finite temperature. Such states play a central role in several fields and applications, such as the study of several quantum statistical mechanics phenomena like thermalization \cite{gogolin2011absence, cramer2012thermalization, riera2012thermalization, gogolin2016equilibration, shirai2020thermalization, chen2021fast, reichental2018thermalization}, out-of-equilibrium thermodynamics \cite{bernard2016conformal, castro2016emergent, brunelli2015out, eisert2015quantum} and open quantum systems \cite{shirai2020thermalization, reichental2018thermalization,shirai2018floquet,scandi2019thermodynamic,lange2018time,rivas2020strong,Poulin_Wocjan_2009}, quantum simulation \cite{childs2018toward}, quantum machine learning \cite{kieferova2017tomography,biamonte2017quantum, bishop2006pattern}, and quantum optimization \cite{kirkpatrick1983optimization, somma2008quantum,krzakala2007gibbs, stilck2021limitations}. Moreover, sampling from well-prepared Gibbs states can be used to tackle several problems, such as semi-definite programming \cite{Brandao_Svore_2017a}, combinatorial optimization problems \cite{kirkpatrick1983optimization, somma2008quantum}, and training quantum Boltzmann machines \cite{Amin_2018, kieferova2017tomography}.

The task of preparing Gibbs states and computing the expectation values of its observables has been proven to be quite cumbersome; at arbitrarily low temperatures, Gibbs state preparation can be considered a QMA-hard problem \cite{watrous2008quantum, Aharonov_Arad_Vidick_2013}. Existing algorithms used for such an implementation can require large overhead in resources and/or calibration even in the simplest of cases, as well as the fact that the implementation might be limited to only a specific set of systems. 

Algorithms that have been proposed consist of implementing the Davies generators which rapidly converge to the Gibbs distribution \cite{Davies_1974, Davies_1976, chen2023quantum,bardet2023rapid, kastoryano2016quantum, Rall_Wang_Wocjan_2022}, the quantum Metropolis algorithm \cite{chen2023quantum, Chiang_Wocjan_2010, Hastings_1970, metropolis1953equation, Temme_2011, Poulin_Wocjan_2009}, and preparing thermal quantum states through quantum imaginary time evolution (QITE)\cite{wang2023critical,Yuan_Endo_Zhao_Li_Benjamin_2019,Tan_2020,Gacon_Zoufal_Carleo_Woerner_2021, Getelina_Gomes_Iadecola_Orth_Yao_2023, McArdle_Jones_Endo_Li_Benjamin_Yuan_2019, motta2020determining, Shtanko_Movassagh_2021, Silva_Taddei_Carrazza_Aolita_2022, Sun_Motta_Tazhigulov_Tan_Chan_Minnich_2021}. Another popular approach amongst near term devices is that of variational quantum algorithms (VQAs) \cite{https://doi.org/10.48550/arxiv.2206.05571, https://doi.org/10.48550/arxiv.2210.16419, Sagastizabal_2021,economou2023role, wu2019variational, chowdhury2020variational, wang2021variational, zhu2020generation, warren2022adaptive,guo2021thermal,ge2016rapid, consiglio2023variational, martyn2019product, Foldager_2022, Premaratne_2020}, where a quantum-classical hybrid approach of minimizing a cost function, using a parameterized quantum circuit (PQC) as a variational ansatz to prepare a Gibbs state is implemented. Despite the prominence of VQAs for Gibbs state preparation, they can require many experimental measurements for each optimization step and may suffer from barren plateaus \cite{coopmans2023predicting, McClean_2018}. To bypass the need to find an ansatz and optimize its parameters could significantly reduce the resources needed for the Gibbs state preparation. 

In certain regimes of locally interacting Hamiltonians on a lattice, with temperatures above any critical point, properties such as the Markov property and uniform clustering property \cite{brandao2019finite} enable classical methods such as linked cluster expansions \cite{oitmaa2006series,sykes1966lattice,tang2013short} and tensor networks to estimate local observables with respect to the underlying high-temperature Gibbs state \cite{tang2013short,kuwahara2021improved}. Additionally, these allow for Gibbs states to be prepared with local constant depth channels via expansions of Quantum Belief Propagation (QBP) \cite{hastings2007quantum,poulin2008belief,bilgin2010coarse,brandao2019finite,kastoryano2016quantum, kim2012perturbative,kato2019quantum}. However, the implementation of these channels in practice may be difficult for near-term devices.     

Inspired by the methods used with QBP and linked cluster expansions, we propose a method based on sampling from a quasi-distribution of local partitions clusters of mixed states, i.e., expanding the full Gibbs state into a sum of products of local “Gibbs-cumulant” type states easier to implement and sample from on quantum hardware. These short-depth circuits come at the cost of a sampling overhead proportional negativity induced by sampling from this pseudo-mixed state. With our method, we outline three different use cases: 1) static observables, 2) dynamical correlation functions,  and 3) its value as a warm start for other quantum algorithms, such as VQAs or QITE.

We will be discussing the development of our algorithm and use cases for it in measuring different observables for prepared Gibbs states on quantum hardware (specifically, IBM quantum hardware) in this manuscript. In Sec.~\ref{meth_sec}, we present the mathematical analysis of our algorithm along with the mathematical description of the different observables we want to measure. In Sec.~\ref{sec_app}, we illustrate the applications we implement our sampling algorithm for. In Sec.~\ref{res_sec}, we specify the expansion of our Gibbs states of choice and present the results of the measurement of the different observables on IBM quantum hardware, followed by a discussion of the overall work in Sec.~\ref{disc_sec} as well as that of future plans.

\section{Methods}\label{meth_sec}
 Our goal in this paper is to sample the Gibbs state as a quasi-distribution over local partitions of mixed states employing a variation of linked cluster expansions. The state preparation circuits for each of these samples may be more feasible to implement in the short term at the cost of a sampling overhead that comes with a magnified variance due to the quasi-distribution of states. We will consider nearest-neighbor lattice Hamiltonians constructed as:
 
 \begin{equation}
     H=\sum_{\langle i,j\rangle}v_{i,j}
 \end{equation}
 where $v_{i,j}$ acts on vertices that share a common nearest-neighbor edge. We also use the notation $H_n$ to define the Hamiltonian acting on $n$-sites. The Gibbs state on $n$-sites is defined as:
 \begin{equation}
     \rho_n(\beta)=\frac{e^{-\beta H_n}}{Tr[e^{\beta H_n}]}
 \end{equation}
 where $\beta$ is the inverse temperature. For short-hand, we will also define $\rho_n=\rho_n(\beta)$.
\subsection{Linked cluster expansion}
Let us start by defining the following cluster cumulants:
\begin{equation}
    \Delta_1 =\rho_1
\end{equation}
\begin{equation}
    \Delta_2=\rho_2-\rho_1^{\otimes 2}
\end{equation}
\begin{equation}
    \Delta_3=\rho_3 -\rho_2\otimes\rho_1-\rho_1\otimes\rho_2+\rho_1^{\otimes 3}
\end{equation}
\begin{equation}
    \Delta_n=\rho_n-(\text{bi-paritions})+(\text{tri-paritions})-...
\end{equation}
where $\rho_n$ is a Gibbs state prepared on $n$-contiguous sites. In two dimensions or greater, there will be multiple distinct lattice topologies for $\Delta_n>2$ so   $\Delta_{T(n)}$ is used to distinguish each distinct lattice topology. If explicitly working in 1-D then we will drop the $T(m)$ notation. An important property to note is that $Tr[\Delta_1]=1$ and $Tr[\Delta_{T(m)}]=0$ if $m>1$. 

These cluster cumulants can now be used to define the full Gibbs state on N-sites as:
\begin{equation}\label{eq:cluster}
   \begin{split}
       &\rho_N=\sum_{j=1}^{N}\sum_{\substack{m_1,m_2,...m_j=1\\m_1+m_2+...+m_j=N}}^{N}\\
       &\Delta_{T(m_1)}\otimes\Delta_{T(m_2)}\otimes...\otimes\Delta_{T(m_j)}.
   \end{split}
\end{equation}
where the sum represents all possible contiguous j-partitions of the lattice.
\subsection{Cluster Sampling}
If the aim is to sample in the high-temperature regime, $\beta\leq\beta_c$, it is expected that $||\Delta_{T(m)}||_1$ ($||\bullet||_1=$ Hilbert-Schmidt norm) will decay exponentially with $m$, meaning that we can exclude terms in the series that include $\Delta_{T(m)}$ with $m$ greater than some cut-off size $m_c$. Keep in mind that $\Delta_{T(m)}$ is not a mixed state since it has zero trace but can be expressed as a linear combination of 2 mixed states:
\begin{equation}
    \begin{split}
         \Delta_{T(m)}&=a\big(\sigma_+-\sigma_-\big) ,\ a\geq0\\
         & ||\Delta_{T(m)}||_1=a
    \end{split}
\end{equation}
By defining the negativity of the pseudo-mixed state as:
\begin{equation}
   \begin{split}
        \lambda=1+\sum_{j=1}^{N-1}\sum_{\substack{m_1,...m_j,\\m_1+...+m_j=N}}||\Delta_{T(m_1)}||_1...||\Delta_{T(m_j)}||_1
   \end{split}
\end{equation}
we could now sample states from each term in the series with probability
\begin{equation}
    \frac{||\Delta_{T(m_1)}||_1...||\Delta_{T(m_j)}||_1}{\lambda}
\end{equation}
with a cost of the variance of any observable being magnified by a factor of $\lambda^2$. 

\subsection{Refined Cluster Sampling}
We instead take a different approach. Assume the goal is to sample from the Gibbs state on $N$-sites but given hardware or depth constraints we only have access to a method to sample Gibbs states up to $M$-sites with $M<N$ and assume $N$ is some integer multiple of N, i.e. $N=j\times M$. The most naive approximation would be to sample from:
\begin{equation}
    \rho_N^{\prime}=\rho_M^{\otimes j}
\end{equation}
which would have error of $\epsilon=\rho_N-\rho^{\prime}_N$. The terms that compose $\epsilon$ consist of all the terms from Eq.~\eqref{eq:cluster} that intersect with the boundary of the partition given by $\rho_N^{\prime}$. It is now possible to add correction terms to $\rho_N^{\prime}$ such as $\rho_{M-1}\otimes\Delta_2\otimes\rho_{M-1}\otimes\rho_M^{\otimes(j-2)}$, $\rho_{M}^{\otimes (j-2)}\otimes\rho_{M-2}\otimes\Delta_3\otimes\rho_{M-2}$, and so on. Following the same analysis in the previous section one would now be sampling from a pseudo-mixed state with a much smaller negativity.  

As an example, let's work with a 1D chain of $2N$-sites and assume we can use known quantum methods to sample from Gibbs states up to $N$-sites with our given limitations. The series that would be sampled would now look like:
\begin{equation}
    \begin{split}
        \rho_{2N}&=\rho_N\otimes\rho_{N}+\rho_{N-1}\otimes\Delta_2\otimes\rho_{N-1}\\
        &+\rho_{N-2}\otimes\Delta_3\otimes\rho_{N-1}+\rho_{N-1}\otimes\Delta_3\otimes\rho_{N-2}\\
        &+\rho_{N-3}\otimes\Delta_4\otimes\rho_{N-1}+\rho_{N-1}\otimes\Delta_4\otimes\rho_{N-3}\\
        &+\rho_{N-2}\otimes\Delta_4\otimes\rho_{N-2}+...
    \end{split}.
\end{equation}

Now let's assume that we truncate our series at $\Delta_m$ with $m>4$. Here, one can pick any of their favorite quantum Gibbs state sampling algorithms for $\rho_N,\rho_{N-1},\rho_{N-2}$, and $\rho_{N-3}$. It is also classically easy to diagonalize $\Delta_2,\Delta_3$, and $\Delta_4$ as well as classically transpile circuits that prepare all the eigenstates that are sampled from each of those operators.  This now results in a negativity given by:
\begin{equation}
    \lambda^{\prime}=1+||\Delta_2||_1+2||\Delta_3||_1+3||\Delta_4||_1
\end{equation}
where there will be a variance magnification of $(\lambda^{\prime})^2$ for any observable, $O$, and a truncation bias of:
\begin{equation}
    \text{bias}<||O||_2\sum_{m=5}^N (m-1)||\Delta_m||_1
\end{equation}

In general, each term in the series that we sample from is made up of products of $\rho_j$'s and $\Delta_{T(k)}$'s with spectral decomposition of:
\begin{equation}
    \begin{split}
        &\rho_j=\sum_{n=0}^{2^j-1}\frac{e^{-\beta E^{(j)}_n}}{Z_j}|n^{(j)}\rangle\langle n^{(j)}|\\
        &\Delta_{T(k)}=\sum_{m=1}^{2^k-1}\gamma^{(k)}_m|m^{(k)}\rangle\langle m^{(k)}|
    \end{split}
\end{equation}
so the the state $|n^{(j)}\rangle\otimes|m^{(k)}\rangle\otimes|n^{(l)}\rangle$ would be sampled from the term $\rho_j\otimes\Delta_{T(k)}\otimes\rho_l$ probability:
\begin{equation}
    \begin{split}
        &p\Big(|n^{(j)}\rangle\otimes|m^{(k)}\rangle\otimes|n^{(l)}\rangle\Big)\\
    &= \frac{||\Delta_{T(k)}||}{\lambda}\frac{e^{-\beta E^{(j)}_n}}{Z_j}\frac{e^{-\beta E^{(l)}_n}}{Z_l}|\gamma_m^{(k)}|.
    \end{split}
\end{equation}

\begin{figure}[htp!]
    
    \centering
    
    \includegraphics[width=\columnwidth]{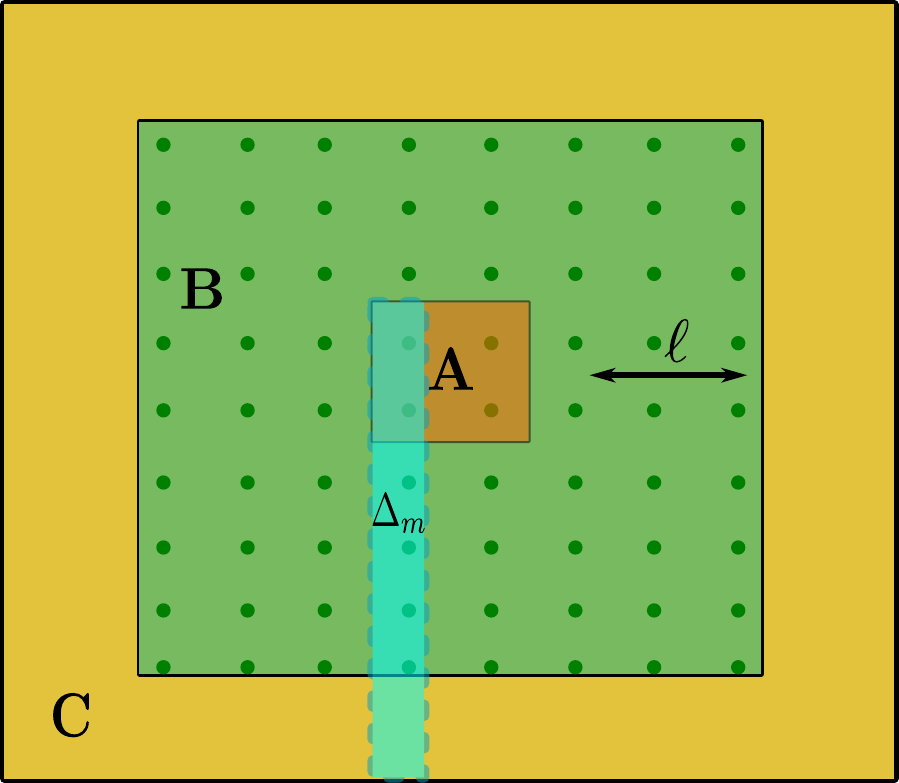}
    
    \caption{\textbf{Lattice decomposition:} A lattice decomposition in terms of the disjoint subset $ABC = \Lambda$, where $\Lambda$ is the full-lattice system. $Tr[\rho_{ABC}O_A]$ is approximated by $Tr[\rho_{AB}O_A]$ with corrections given by $Tr[\Delta_mO_A]$ where $\Delta_m$ represents all the correlators with support on all three regions, A, B, and C and $O_A$ is an operator supported on A. }
    \label{rhoABC}
\end{figure} 

\section{Applications}\label{sec_app}

While the application of our sampling procedure can be used in any situation, we are more interested in its utility in calculating response functions, non-local observables, or serving as a warm-start for other quantum algorithms vs calculating static local observables. In the first few cases, it makes sense to use the procedure we have laid out in the previous section. When it comes to local observables we outline a different procedure.

In this scenario, we have a partition of our ideal lattice $\Lambda$ into 3 subsystems, $ABC=\Lambda$, such that $B$ shields $A$ from $C$ as shown in Fig.~\ref{rhoABC} . We would like to calculate $\Tr[\rho_{\Lambda}O_A]$, where $O_A$ is supported on region $A$ but can only sample from the Gibbs state $\rho_{AB}$. 

Instead of running the full sampling algorithm from the previous section, it makes sense to only calculate $\Tr[\rho_{AB}O_A]$ with the quantum computer. We can then add classically calculated corrections which will consist of $\Delta_m$'s that have support on all three regions. We can classically add these corrections to $\Tr[\rho_{AB}O_A]$ up to sizes that are classically tractable. This equates to running standard cluster expansion algorithms that include clusters that cannot be classically computed. 

We now illustrate the different applications we use our sampling algorithm for in this manuscript. 

\subsection{Dynamical spin-spin correlation functions}
In the case of $T = 0$, the dynamical spin-spin correlation function can be written as:
\begin{equation}
\begin{split}\label{eq:CorrelationFunc}
    C^{\alpha \gamma} _{i,j} (t)  &:=  \langle  s^{\alpha} _i (t) s^{\gamma} _j (0)  \rangle_{0} \\&=  \sum_{p} \bra{0}s^\alpha_{i}\ket{p}\bra{p}s^\gamma_{j}\ket{0} e^{-iE_{p} t} \, .
\end{split}
\end{equation}
For the case of $\alpha = \gamma = z$ and measuring the correlation function of the Gibbs state, we can write out the $Z_iZ_j$ correlation function as follows:
\begin{equation}
    \langle Z_i (t) Z_j (0)\rangle_\beta = Tr[\rho(\beta, H) Z_i (t) Z_j (0)] \, ,
\end{equation}
where $i,j$ are the indices of the spins.

In our work in Ref.~\cite{eassa2023high}, we work on measuring $T = 0$ correlation functions for dimers. We use similar methods to measure the $ZZ$ correlation functions of the Gibbs states, with the local Gibbs clusters set as the initial state in this case and the correlation functions being weighed accordingly (further details in Sec.~\ref{sec_rho4}). With this, we calculate the transverse dynamical structure factor
\begin{equation}\label{DFS}
    \begin{split}
        S^{Z,Z}(\mathbf{Q},\omega) =  \frac{1}{2\pi\hbar} \int^{\infty}_{-\infty} dt\hspace{3pt} e^{-i \omega t } &\\
        \times \frac{1}{N} \sum_{i,j} \langle Z_i (t) Z_j (0)\rangle_\beta  e^{-i\mathbf{Q}\cdot \mathbf{R}_{ij}} .
    \end{split}
\end{equation}

\subsection{Specific heat}
We also aim to measure the specific heat of the Gibbs states. We define it as so:
\begin{equation}\label{specheq}
    \begin{aligned}
        \begin{split}
            C_\nu &= \frac{1}{\abs{\Lambda}} \frac{\partial \langle H \rangle}{\partial T} \\
            &= \frac{1}{\abs{\Lambda}T^2}[\langle H^2 \rangle - \langle H \rangle^2] \,,
        \end{split}
    \end{aligned}
\end{equation}
where $\abs{\Lambda} = N_1 \times \cdots \times N_D$ is the volume of the lattice $\Lambda$ and $D$ is the dimensionality. Finding the expectation value of the Hamiltonian, $\langle H \rangle$, is equivalent to finding the expectation value of each of the terms of the Hamiltonian and weighing them accordingly. 

\section{Results}\label{res_sec}
\subsection{$\mathbf{\rho_4}$ Gibbs state results}\label{sec_rho4}
\begin{figure}[ht!]
    
    \centering
    
    \includegraphics[width=\columnwidth]{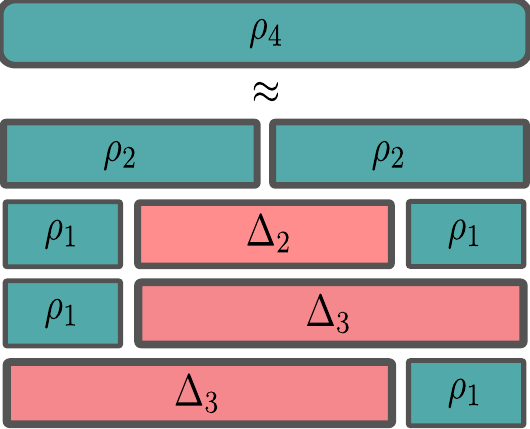}
    
    \caption{$\boldsymbol{\rho_4}$ \textbf{approximation}: Approximation of $\rho_4$ with error given by $||\rho_4-\rho_4^{\prime}||_1=||\Delta_4||_1$.}
    \label{rho4}
\end{figure}

We approximate the global Gibbs state $\rho_4$ of the 4-spin chain as follows:
\begin{equation}\label{rho4ap}
    \rho_4 \approx \rho_2 \otimes \rho_2 + \rho_1 \otimes \Delta_2 \otimes \rho_1 + \Delta_3 \otimes \rho_1 + \rho_1 \otimes \Delta_3 \, ,
\end{equation}
which is illustrated in Fig.~\ref{rho4}. 

\begin{figure}[ht!]
    
    \centering
    
    \includegraphics[width=1.0\columnwidth]{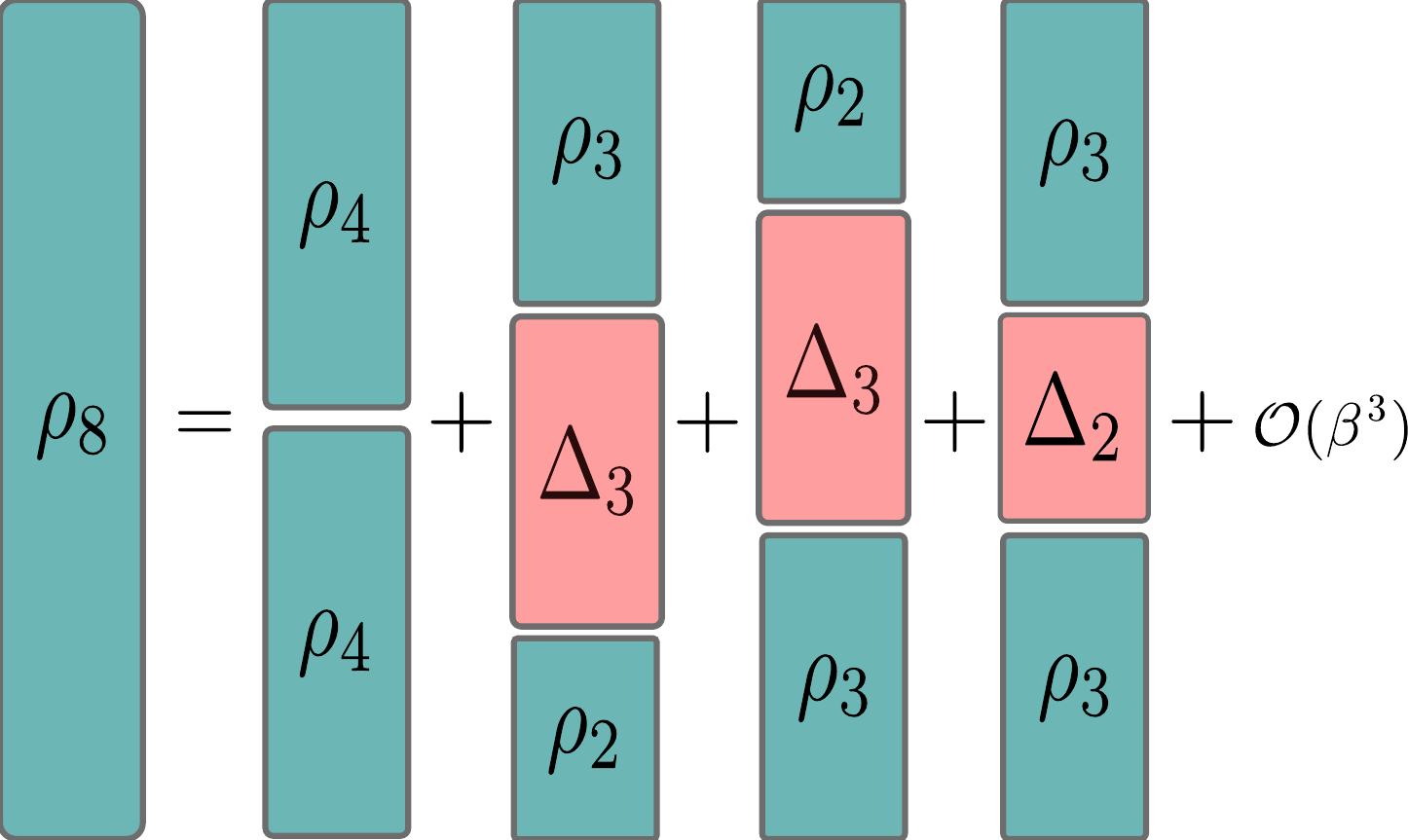}
    
    \caption{$\boldsymbol{\rho_8}$ \textbf{approximation}: Approximation of $\rho_8$ with error bounded by $||\rho_8-\rho_8^{\prime}||_1\leq 3||\Delta_4||_1+4||\Delta_5||_1+3||\Delta_6||_1+2||\Delta_7||_1+||\Delta_8||_1$.}
    \label{rho8}
\end{figure} 
Hence, in order to calculate a thermodynamic observable of $\rho_4$, we calculate the following:\\
\begin{equation}\label{trr4}
    \Tr[\rho_4 O] \approx Tr[\rho_2 \otimes \rho_2 O] + Tr[\rho_1 \otimes\Delta_2 \otimes \rho_1] + \cdots
\end{equation}
and so on.

For our results, we prepare Gibbs states with the Hamiltonian $H$ being set as the 1D XY model Hamiltonian, 
\begin{equation}
    H_{XY} = \sum_{i = 1}^{n} X_i X_{i+1} + Y_i Y_{i+1} \,.
\end{equation}
We define $H_2$ for the XY dimer,

\begin{equation}
    H_{2} = X_1X_2 + Y_1Y_2 \, .
\end{equation}

We can explicitly write the first two terms of the approximation in Eq.~\eqref{trr4} in terms of the eigenstate basis of $H_2$ and the computational basis,
\begin{equation}
    \begin{aligned}
        \Tr[\rho_2\otimes\rho_2 O] &= \sum^{3}_{n = 0}\sum^{3}_{m = 0} p_n p_m \bra{nm}O\ket{nm} \\
        \Tr[\rho_1\otimes\Delta_2\otimes\rho_1 O] &= \sum^{1}_{x=0}\sum^{3}_{n = 0}\sum^{1}_{y = 0} \frac{1}{4}(p_n -\frac{1}{4}) \bra{xny}O
        \ket{xny} \,,
    \end{aligned}
\end{equation}
where $n,m\in$ the eigenstate basis of $H_2$,  $x,y\in$ the computational basis, and the probabilities $p_0, p_1, p_2, p_3$ can be written as:
\begin{equation}
    \begin{cases}
        \begin{aligned}
            p_0 &= \frac{e^{2\beta}}{2(1+\cosh{(2\beta)})}\\
            p_1 &= p_2 = \frac{1}{2(1+\cosh{(2\beta)})}\\
            p_3 &= \frac{e^{-2\beta}}{2(1+\cosh{(2\beta)})}
        \end{aligned}
    \end{cases}
\end{equation} 
with $Z_2 = 2 + e^{2\beta} +e^{-2\beta} = 2(1 + \cosh{(2\beta)})$ being the partition function of the system. As for the $\Delta_3$ terms, each simulation result is weighted by the eigenvalue of the corresponding eigenvector the state maps to (which is essentially what is done for the first and second-order terms as well).

\begin{figure}[ht!]
    
    \centering
    
    \includegraphics[width=1\columnwidth]{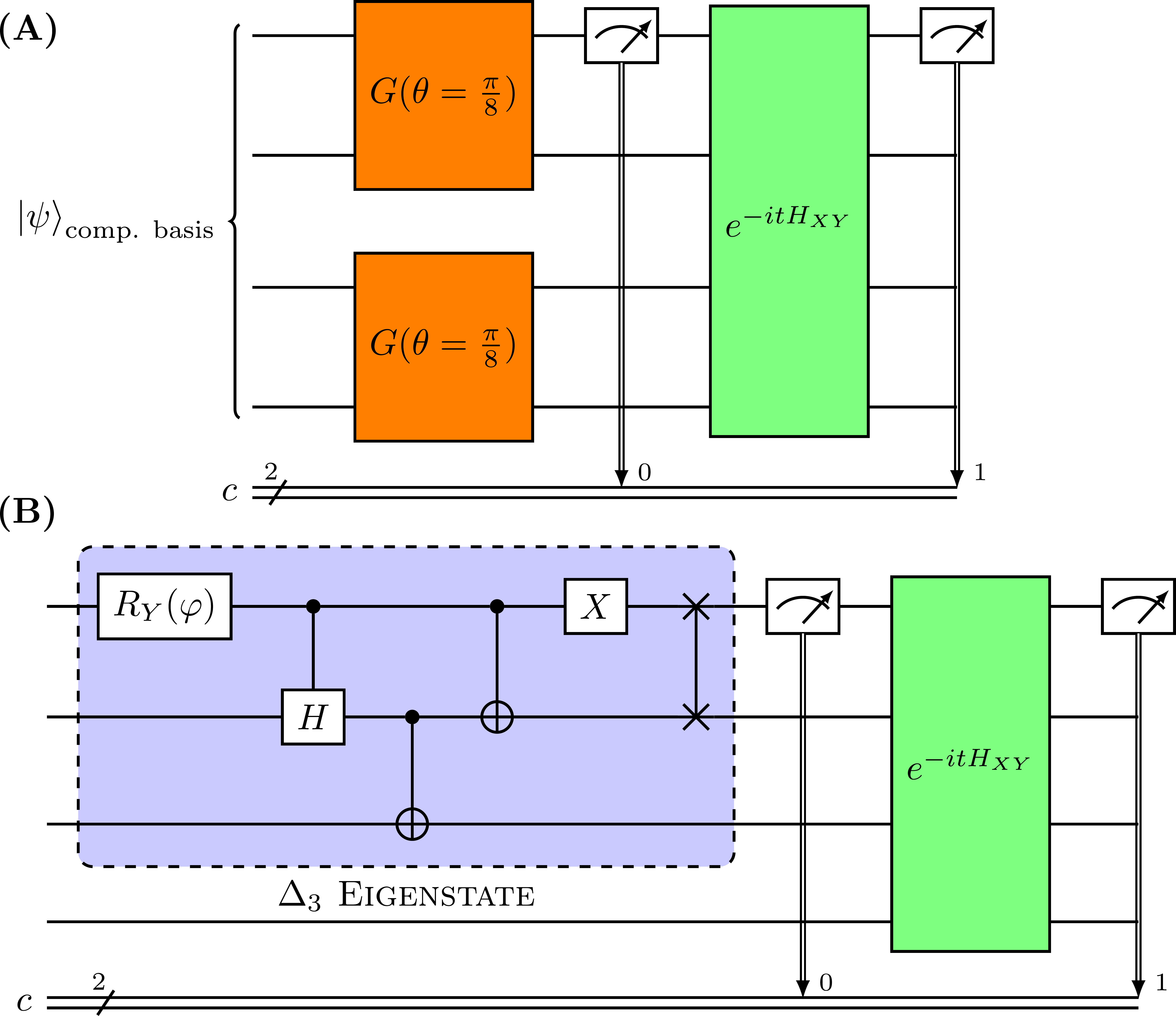}
    
    \caption{\textbf{Sampling circuit examples}: \textbf{(A)} An example of one of the circuits used for the sampling of the $\rho_2\otimes\rho_2$ term of the approximation of $\rho_4$ and measuring the $Z_1 Z_1$ correlation function. The computational basis can be implemented by applying different combinations of $X$ gates, amounting to 16 different combinations representing the 16 computational bases of 4-spin 1/2 systems. \textbf{(B)} An example of one of the circuits used for the sampling of the $\Delta_3\otimes\rho_1$ term of the approximation of $\rho_4$ and measuring the $Z_1 Z_1$ correlation function. The correlation function obtained from this circuit is weighted by the eigenvalue of the $\Delta_3 \otimes \rho_1$ state implemented as the initial state. The direct measurement scheme is employed in all cases. }
    \label{corr_circ}
\end{figure}

Similar to what we have done for our work in Ref.~\cite{eassa2023high}, we implement the direct measurement scheme to measure the $ZZ$ correlation functions as we set the initial state as the different local Gibbs clusters we prepare to be able to approximately measure the $ZZ$ correlation functions of the full $\rho_4$ Gibbs state. All of our hardware results were run on Qiskit Runtime \cite{qiskit} using the Estimator primitive, where we applied twirled readout error extinction (T-REx) \cite{van2022model}.  In Fig.~\ref{corr_circ}, we illustrate examples of the different circuits implemented to measure the $ZZ$ correlation functions of \textbf{(A)} the $\rho_2 \otimes \rho_2$ term and \textbf{(B)} the $\Delta_3 \otimes \rho_1$ term in the expansion.

\begin{figure}[ht!]
    
    \centering
    
    \includegraphics[width=1\columnwidth]{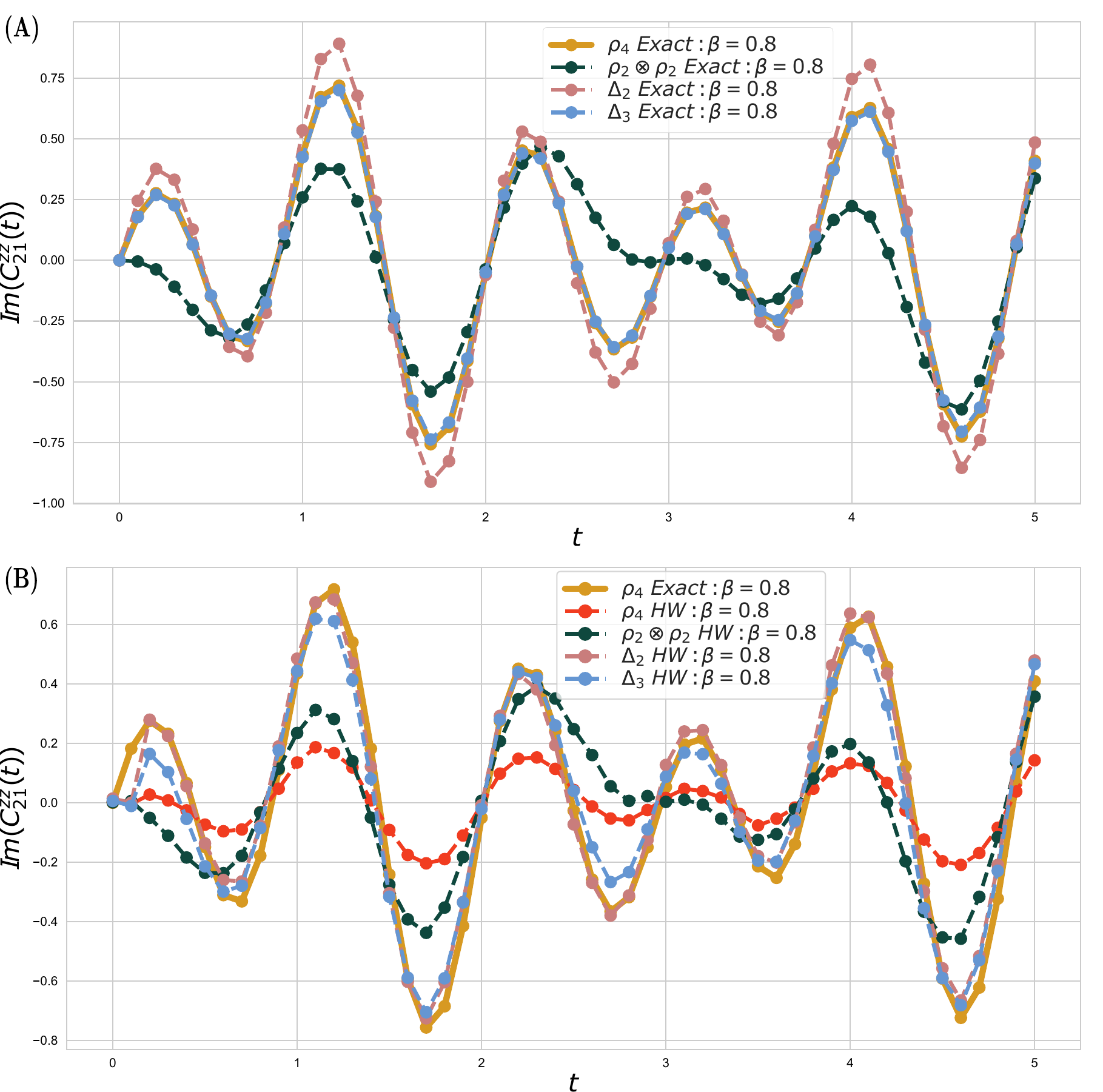}
    
    \caption{\textbf{Comparison of accuracy of the approximation vs. the full Gibbs state:} Results of measuring the imaginary part of the correlation function $C^{zz}_{2,1}(t)$ over time for the $\rho_4$ Gibbs state are presented. The Hamiltonian employed is the XY model at the inverse temperature $\beta = 0.8$. Results were obtained using \textit{ibm\_algiers}, with the qubits' initial layout of [2,1,4,7] for the direct measurements, along with setting $t = 5$, the number of time steps $= 51$, $dt \approx 0.1$, and the number of shots $= 4000$. In panel \textbf{(A)}, the analytical result of each order of the approximation as illustrated in Eq.~\eqref{rho4ap} is presented, as well as the analytical result of simulating the full Gibbs state (yellow). In panel \textbf{(B)}, the corresponding quantum hardware results are presented. The implementation of the full Gibbs state simulation on hardware is shown in red. In both panels, the first order is in dark green, the second order is in pink, and the third order is in blue. We observe that all orders of the approximation we have developed with our sampling algorithm perform significantly better than simulating the full Gibbs state on hardware. It should be noted that the analytical result of the second order of the approximation in \textbf{(A)} demonstrates an overshoot, especially compared to the third order. We can see there is a general amplitude damping caused by the noisy hardware with the hardware results in \textbf{(B)}. Hence, it would seem that the second-order result on hardware is more accurate than the third, but it is just a coincidence for this correlation function. In general, the third-order results on hardware and analytically are more accurate. }
    \label{rh04_im}
\end{figure}

\begin{figure*}[ht!]
    
    \centering

    \includegraphics[width=1\textwidth]{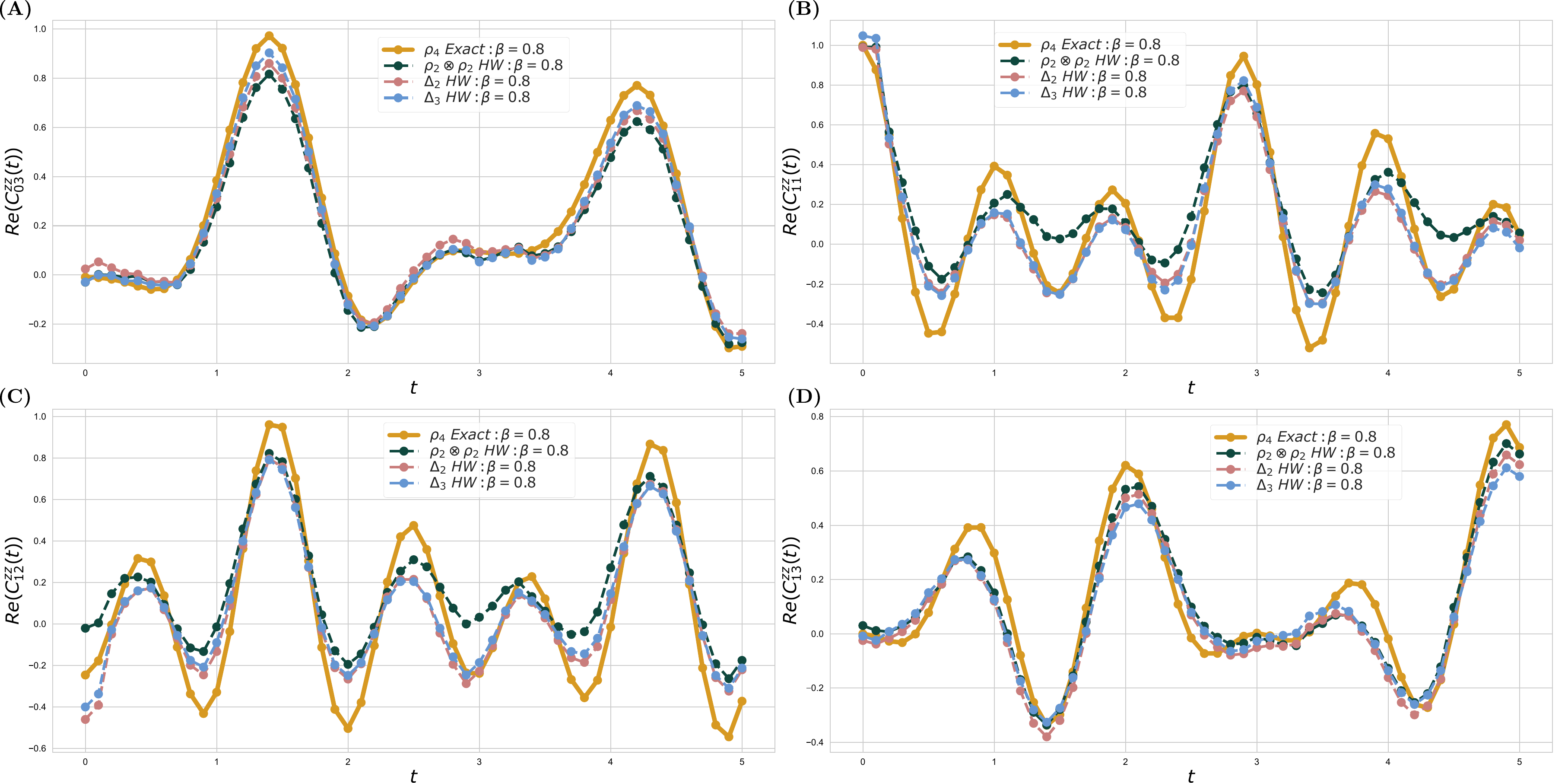}
    
    \caption{\textbf{Different correlation functions on hardware:} Results of measuring the real part of the correlation function \textbf{(A)} $C^{zz}_{1,1}(t)$, \textbf{(B)} $C^{zz}_{1,2}(t)$, \textbf{(C)} $C^{zz}_{1,3}(t)$, and \textbf{(D)} $C^{zz}_{0,3}(t)$ over time for the approximation of the $\rho_4$ Gibbs state are presented. The Hamiltonian employed is the XY model at the inverse temperature $\beta = 0.8$. Results were obtained using \textit{ibm\_algiers}, with the qubits' initial layout of [2,1,4,7] for the direct measurements, along with setting $t = 5$, the number of time steps $= 51$, $dt \approx 0.1$, and the number of shots $= 4000$. The analytical result of the full $\rho_4$ Gibbs state (yellow) is plotted with the hardware simulation result of each order of the approximation against time $t$.  The first (dark green), second (pink), and third (blue) order hardware results are all plotted. An increase in accuracy is observed with each order of the approximation.}
    \label{rho4_panels}
\end{figure*}

We first demonstrate how well our sampling algorithm performs compared to preparing the full Gibbs state when implemented on quantum hardware. In Fig.~\ref{rh04_im}, we show the results of simulating the imaginary part of the correlation function $C^{zz}_{2,1}(t)$ over time $t$ at $\beta = 0.8$, having the system initialized with the different orders of the approximation in Eq.~\eqref{rho4ap}, along with also showing the full Gibbs state $\rho_4$ simulation of the same correlation function. When implementing on quantum hardware, we see that all of the orders of the approximation perform better than the full Gibbs state compared to the expected analytical result. This goes back to the fact that the approximation terms are all simulated with shorter circuit depths compared to the long circuit depth of the circuit implementation of the full Gibbs state. Hence, we are able to establish that we make better use of our sampling algorithm when wanting to measure the required correlation functions to calculate the dynamical structure factor. In Fig.~\ref{rho4_panels}, we present the hardware simulation results of measuring the real part of the correlation functions $C^{zz}_{1,1}(t)$, $C^{zz}_{1,2}(t)$, $C^{zz}_{1,3}(t)$, and $C^{zz}_{0,3}(t)$. We further show the accuracy of the simulation results of each order of the approximation compared to the analytical result of the full $\rho_4$ Gibbs state simulation.

\begin{figure*}[ht!]
    
    \centering
    
    \includegraphics[width=1\textwidth]{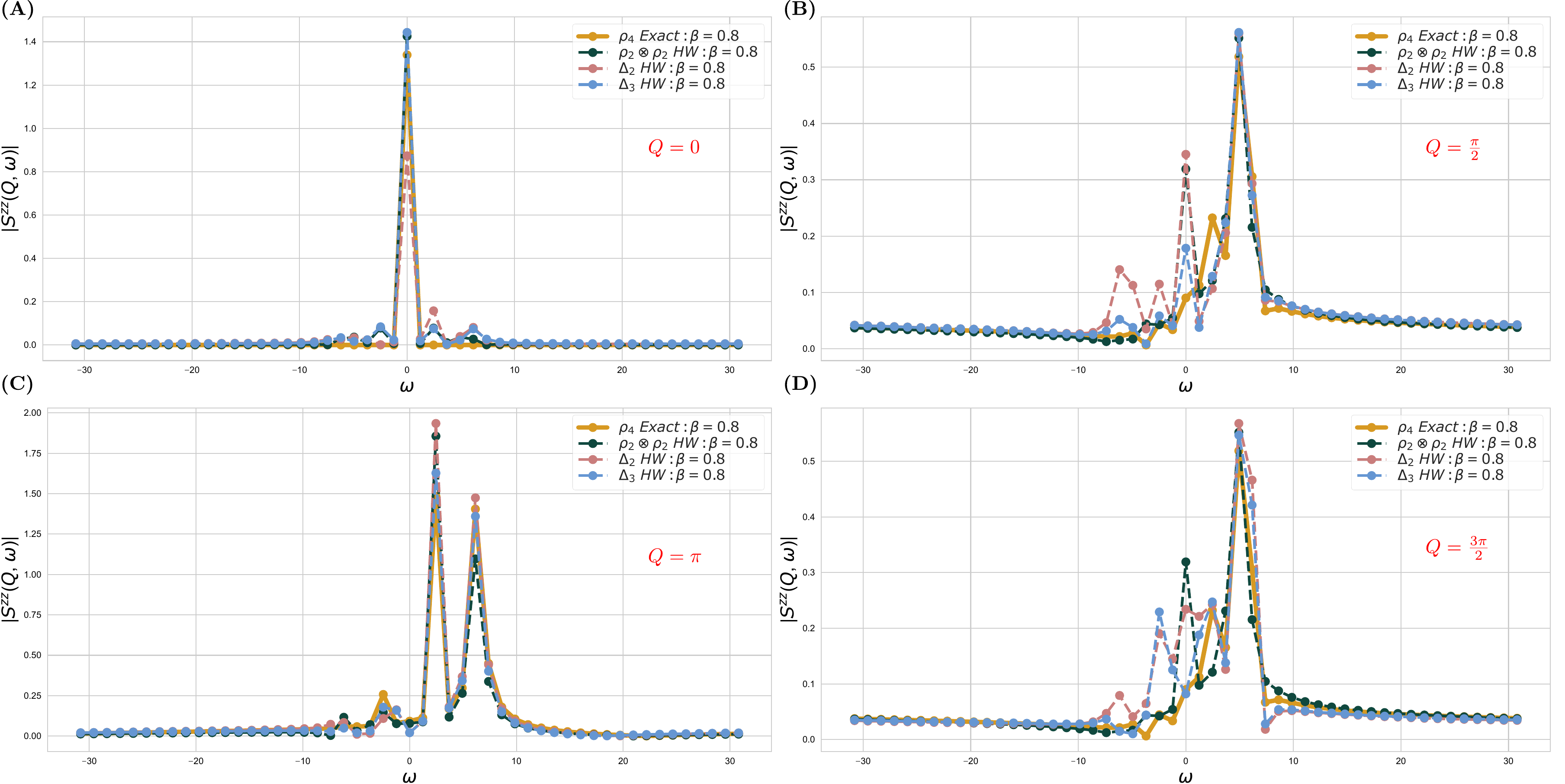}
    
    \caption{\textbf{Dynamical structure factor measurements:} Results of measuring the dynamical structure factor $S(\boldsymbol{Q}, \omega)$ plotted against frequency $\omega$, where the measuring units of $\omega$ is that of $\abs{J}$, for the approximation of the $\rho_4$ Gibbs state are presented. The Hamiltonian employed is the XY model at the inverse temperature $\beta = 0.8$. Results were obtained using \textit{ibm\_algiers}, with the qubits' initial layout of [2,1,4,7] for the direct measurements, along with setting $t = 5$, the number of time steps $= 51$, $dt \approx 0.1$, and the number of shots $= 4000$ to obtain all of the correlation functions needed to calculate the dynamical structure factor (as illustrated in Eq.~\eqref{DFS}). $S(\boldsymbol{Q}, \omega)$ is calculated for the values \textbf{(A)} $\boldsymbol{Q} = 0$, \textbf{(B)} $\boldsymbol{Q} = \frac{\pi}{2}$, \textbf{(C)} $\boldsymbol{Q} = \pi$, and \textbf{(D)} $\boldsymbol{Q} = \frac{3\pi}{2}$. }
    \label{sqw_panels}
\end{figure*}

The goal was to calculate the $zz$ dynamical structure factor $S^{zz}(\boldsymbol{Q},\omega)$. Hence, it was required to measure all of the correlation functions $C^{zz}_{i,j}(t)$,
\begin{equation}
   C^{zz}_{i,j}(t) = \begin{bmatrix}
        C^{zz}_{0,0}(t) &  C^{zz}_{0,1}(t) &  C^{zz}_{0,2}(t) &  C^{zz}_{0,3}(t) \\
        C^{zz}_{1,0}(t) &  C^{zz}_{1,1}(t) &  C^{zz}_{1,2}(t) &  C^{zz}_{1,3}(t) \\
        C^{zz}_{2,0}(t) &  C^{zz}_{2,1}(t) &  C^{zz}_{2,2}(t) &  C^{zz}_{2,3}(t) \\
        C^{zz}_{3,0}(t) &  C^{zz}_{3,1}(t) &  C^{zz}_{3,2}(t) &  C^{zz}_{3,3}(t) 
    \end{bmatrix}\,,
\end{equation}
where all of the spin index combinations are spanned. In Fig.~\ref{sqw_panels}, the results of calculating the dynamical structure factor using all of the correlation functions we measured on quantum hardware are presented in panels \textbf{(A)}, \textbf{(B)}, \textbf{(C)} and \textbf{(D)} for different values of $\boldsymbol{Q}$ ($\boldsymbol{Q} = [0, \frac{\pi}{2}, \pi, \frac{3\pi}{2}$] respectively). We achieve increasingly greater levels of accuracy with the different orders of the approximation.

\begin{figure}[ht!]
    
    \centering
    
    \includegraphics[width=1\columnwidth]{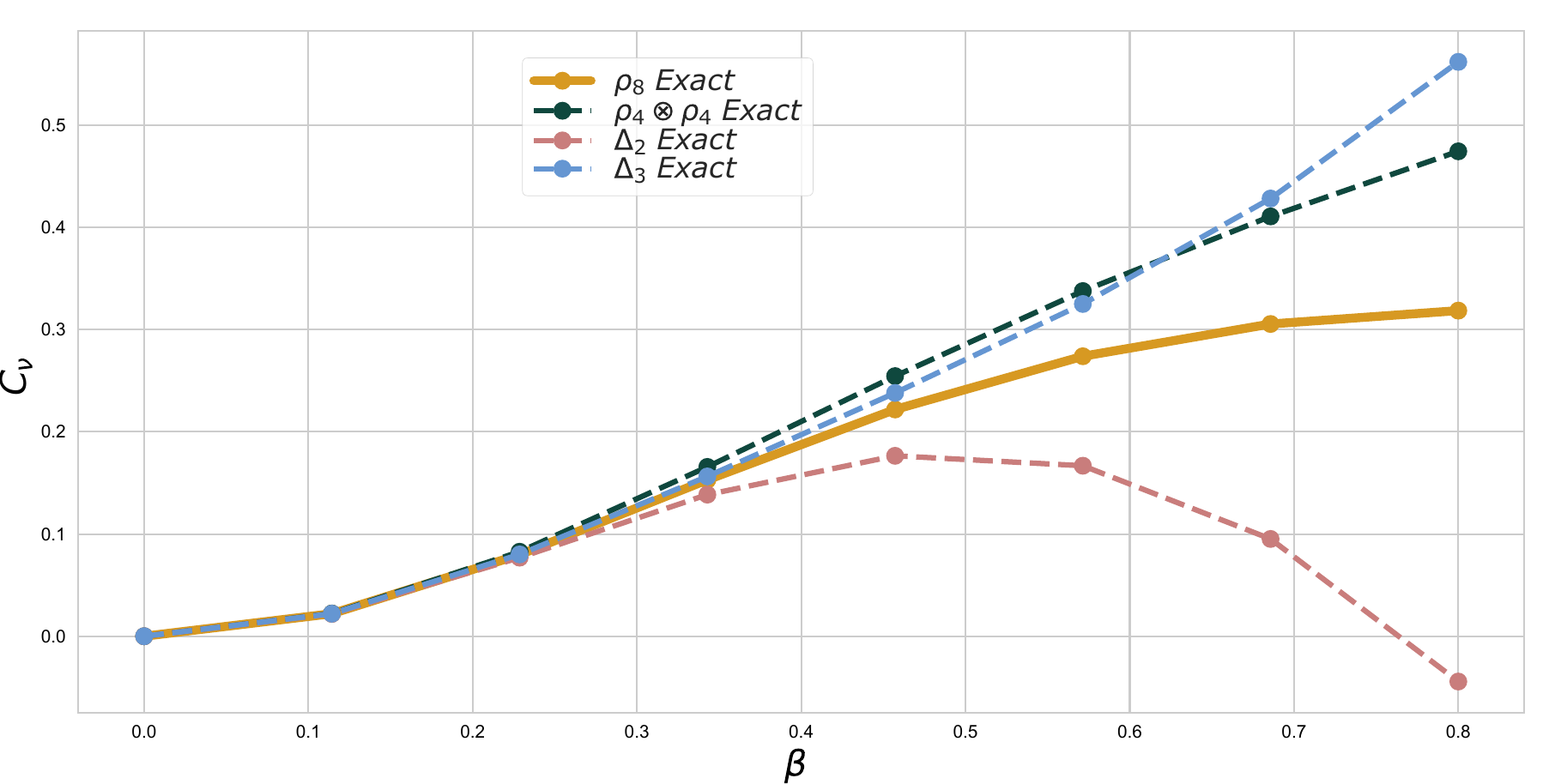}
    
    \caption{\textbf{Analytical results of measuring the specific heat of the $\rho_8$ Gibbs state:} Results of calculating the specific heat of $\rho_8$ approximated by $\rho_4\otimes\rho_4$ with corrections up to $\Delta_3$ plotted against the inverse temperature $\beta$. With larger-sized correction terms, our approximation is predicted to diverge at larger values of $\beta$.}
    \label{cv_anal}
\end{figure} 

\subsection{$\mathbf{\rho_8}$ Gibbs state results}\label{sec_rho8}

We approximate the global Gibbs state $\rho_8$ of the 8-spin chain as follows:
\begin{equation}\label{rho8eqapp}
    \rho_8 \approx \rho_4 \otimes \rho_4 + \rho_3 \otimes \Delta_2 \otimes \rho_3 + \rho_2 \otimes \Delta_3 \otimes \rho_3 + \rho_3 \otimes \Delta_3 \otimes \rho_2 \, ,
\end{equation}
where $\rho_1,\rho_2,\rho_3,\Delta_2,\Delta_3$ all are computed identically to the case of the $\rho_4$ results. $\rho_4$ is written as: 
\begin{equation}
    \rho_4 = \frac{e^{-\beta H_4}}{\Tr[e^{-\beta H_4}]} \,.
\end{equation}
An illustration of the expansion is shown in Fig.~\ref{rho8}.

For $\rho_8$, we measure the specific heat $C_\nu$. The process of measuring it consists of (i) preparing the different expansion terms and (ii) measuring the expectation values $\langle H \rangle$ and $\langle H^2 \rangle$ on the quantum circuits as per the definition of $C_\nu$ in Eq.~\eqref{specheq}, which is equivalent to measuring the expectation values of the individual terms of $H$ and $H^2$. The Hamiltonian in this case is still the 1D XY model. The Givens rotations used in the state preparation of the Gibbs clusters in this case do not change. In Sec.~\ref{spec_heat_sec}, we illustrate an implementation of measuring $\langle H \rangle$ and $\langle H^2 \rangle$ while also cutting down on the required number of circuits to do so.

In Fig.~\ref{cv_anal}, we present the analytical results of calculating the specific heat $C_\nu$ for the different orders of the approximation illustrated in Eq.~\eqref{rho8eqapp}, as well as the full Gibbs state $\rho_8$, for different values of $\beta$. Our algorithm was developed on the basis of being implemented for systems of exponentially decaying correlation length, i.e., systems at higher temperature ranges. Hence, the orders of the approximation start to diverge at larger values of $\beta$ (i.e., lower temperature range). More orders of the approximation can be taken to increase the accuracy of the approximation at larger values of $\beta$.

\section{Discussion}\label{disc_sec}

We have developed an algorithm to prepare Gibbs states via cluster expansions and utilize them for different simulation applications, such as simulating the dynamical structure factors and specific heats of the systems of these states. Our algorithm is more suited to be used for systems with exponentially decaying correlation length, i.e., stable short-range order/higher temperature ranges. The application test cases we presented in this manuscript were all 1D systems with the XY model Hamiltonian, however, the algorithm is not limited to these cases. The next step would be to move onto a 2D system with a different Hamiltonian model. An obstacle we would face is the fact that the number of clusters per approximation order scales with the size of the boundary, meaning that the computational expense would immensely grow. Steps that would greatly help with circumventing such a problem would be identifying the symmetries and equivalent topologies of the different clusters as well as exploring other possible ways to further reduce the sampling overhead. To circumvent the problem of the temperature range spanned by our algorithm, it can be used in tandem and as a warm-start for other algorithms that would be more suited for lower temperature ranges, such as QITE. 

\clearpage
\newpage
\bibliographystyle{ieeetr}
\bibliography{arXivv1.bib}

\section*{Acknowledgements}
All authors and the research as a whole were supported by the Quantum Science Center (QSC), a National Quantum Science Initiative of the Department Of Energy (DOE), managed by Oak Ridge National Laboratory (ORNL). We acknowledge the use of IBM Quantum services for this work. This research used resources from the Oak Ridge Leadership Computing Facility, which is a DOE Office of Science User Facility supported under Contract No. DE-AC05-00OR22725. 
\section*{Author Contributions}
JC conceived the theoretical basis of the project. The test cases of the applications presented were discussed amongst NME, JC, and AB. NME performed all of the simulations and data analysis, with input from JC. JC formulated the Givens rotation decomposition of the time evolution operator. MMM helped with earlier simulator results in the project. NME produced the first draft with input from JC. JC and NME worked on finishing the final draft of the manuscript.
\section*{Competing Interests}
Authors declare that they have no competing interests.
\section*{Data and Materials Availability}
Data is available upon request.
\clearpage
\onecolumngrid
\section{Supplementary materials}
\beginsupplement

\subsection{Linked-cluster expansions}
We find a strong basis for our work in linked-cluster expansions (LCE). The idea behind LCE \cite{oitmaa2006series,sykes1966lattice,tang2013short} is that the value of any extensive property (examples would be the logarithm of the partition function, internal energy, Gibbs states, etc.) of a lattice model can be computed per lattice site $P(\mathcal{L})/N$ in the thermodynamic limit in terms of a sum of contributions from all clusters c that can be embedded on the lattice
\begin{equation}\label{clust}
    P(\mathcal{L})/N = \sum_c L(c) \times W_P (c) \, , 
\end{equation}
where $L(c)$ is the multiplicity of $c$, namely the number of ways per site in which the cluster $c$ can be embedded on the lattice, and $W_P(c)$ is the weight of that cluster for the property $P$. $W_P(c)$ is defined according to the inclusion-exclusion principle:
\begin{equation} \label{in_ex}
    W_P(c) = P(c) - \sum_{s\subset c} W_P(s) \, , 
\end{equation}
where 
\begin{equation}
    P(c) = \frac{\text{Tr}[P(c) e^{-\beta H_c}]}{\text{Tr}[e^{-\beta H_c}]}
\end{equation}
is the property calculated for the finite cluster $c$ and the sum on $s$ runs over all the subclusters of $c$. $H_c$ is the Hamiltonian of cluster $c$. 

Because of the inclusion-exclusion principle in Eq.~\eqref{in_ex}, the weight of every cluster contains only the contribution to the property that results form the correlations that involve all the sites in the cluster, and in a unique fashion according to its specific geometry. At low temperature, when the correlations grow beyond the size of the largest clusters considered in the series, the results diverge as we lose the contributions of clusters in higher orders of the expansion. This usually occurs near or at zero temperature for most 2D quantum models of interest, e.g., the nearest-neighbor antiferromagnetic (AF) Heisenberg model on a bipartite lattice. 

It is important to note that LCEs can still be quite computationally demanding. Several steps need to be taken in the implementation of LCEs, such as (i) generating all the linked clusters that can be embedded on the lattice, (ii) identifying their symmetries and topologies to compute the multiplicities (which also reduces the computational expense of the implementation), (iii) identifying the subclusters to calculate the weights, and (iv) calculating the property of each cluster and performing the sums. The number of the embedded clusters and subclusters grow exponentially with increasing the order of the expansion. 

In the LCEs, the only clusters included are \textit{connected} as it can be proven that for all disconnected clusters, the weight vanishes as $P(c)$ can be written as the sum of its parts. For example, given that we have two disconnected subclusters $c_1$ and $c_2$. We can then write out $P(c)$ as:
\begin{equation}
    P(c) = P(c_1) + P(c_2) .
\end{equation}
However, $c_1$ and $c_2$ are themselves subclusters of $c$. Hence,
\begin{equation}
    \begin{aligned}
        \begin{split}
            W_P(c) &= P(c) - \sum_{s\subset c} W_P(s) \\
            &= P(c) - [W_P(c_1) + \sum_{s\subset c_1} W_P(s)] \\
            &-[W_P(c_2) + \sum_{s\subset c_2} W_P(s)] \\
            &= P(c) - P(c_1) - P(c_2) = 0 .
        \end{split}
    \end{aligned}
\end{equation}
Hence, the name linked-cluster expansions.

The clusters in Eq.~\eqref{clust} are usually grouped together based on common characteristics to form different orders of the expansion \cite{rigol2006numerical,rigol2007numerical}. In the case of LCEs, one has the freedom to devise an expansion with a certain building block for generating the clusters in different orders, depending on what is best for the model of choice. Examples of building blocks that can be used are the sites or bonds in the lattice. 

\subsection{Givens rotations and the Jacobi transformation}\label{givens_sec}
Givens operators represent rotations in a two-dimensional subspace while leaving all other dimensions invariant \cite{givens1958computation, frerix2019approximating}. The matrix representation of the Givens rotation operator in the $(i,j)$ plane by an angle $\theta$ can be written as
\begin{equation}
    G(i,j,\theta) = \begin{bmatrix}
     1 &\cdots&0&\cdots&0&\cdots&0 \\
     \vdots &\ddots&\vdots& &\vdots& &\vdots \\
     0 &\cdots&\cos{(\theta)}&\cdots&\sin{(\theta)}&\cdots&0 \\
     \vdots & &\vdots& \ddots &\vdots& &\vdots \\
      0 &\cdots&-\sin{(\theta)}&\cdots&\cos{(\theta)}&\cdots&0 \\
      \vdots & &\vdots&  &\vdots& \ddots &\vdots \\
      0 & \cdots & 0 & \cdots &0& \cdots &1 
    \end{bmatrix}\,,
\end{equation}
where the trignometric expressions appear in the $i$-th and $j$-th rows and columns respectively. It can be shown that any $n \times n$ orthogonal matrix can be written as a product of at most $\frac{n(n-1)}{2}$ Givens rotations \cite{aronov2012discrete, frerix2019approximating}. 

For our results, we prepare Gibbs states with the Hamiltonian $H$ being set as the 1D XY model Hamiltonian, 
\begin{equation}
    H_{XY} = \sum_{i = 1}^{n} X_i X_{i+1} + Y_i Y_{i+1} \,.
\end{equation}
We begin with an XY dimer:

\begin{equation}
    H_{2} = X_1X_2 + Y_1Y_2 \,,
\end{equation}
where
\begin{equation}\label{XYdimer}
  \begin{cases}
    \begin{aligned}
      &H_{2}\ket{00}\hspace{8pt} = 0\ket{00} \\
      &H_{2}\ket{11}\hspace{8pt} = 0\ket{11} \\
      &H_{2}\frac{1}{\sqrt{2}}[\ket{01} + \ket{10}] = \pm 2 \frac{1}{\sqrt{2}}[\ket{01}\pm\ket{10}]
    \end{aligned}
  \end{cases} 
\end{equation}

\begin{figure*}[ht!]
    
    \centering
    
    \includegraphics[width=\textwidth]{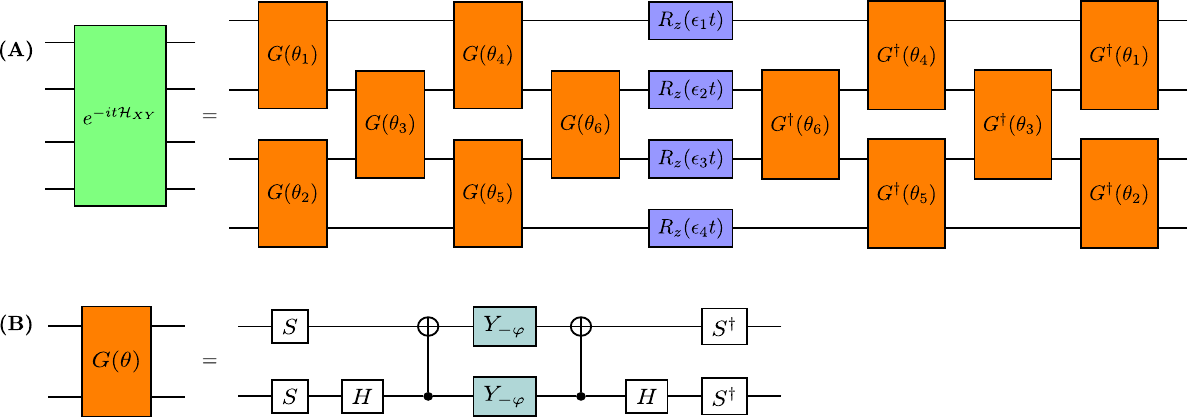}
    
    \caption{\textbf{Givens rotations and time evolution operator:} \textbf{(A)} The decomposition of the time evolution operator via a series of Givens rotations. \textbf{(B)} The circuit representation of the Givens rotation operator described in Eq.~\eqref{givensrot}.}
    \label{ham_diag_fig}
\end{figure*}

We define the Givens rotation operator (see Sec.~\ref{givens_sec}) in terms of the Pauli operators as so:
\begin{equation}\label{givensrot}
    \begin{aligned}
        G_{\theta} = e^{-i\theta(XY - YX)} &= \cos^2{(\theta)} \id +\sin^2{(\theta)}ZZ \\
        & - i\sin{(\theta)}\cos{(\theta)}[XY - YX] \,,
    \end{aligned}
\end{equation}
where 
\begin{equation}\label{givens}
  \begin{cases}
    \begin{aligned}
      &G_{\theta}\ket{00}\hspace{8pt} = \ket{00} \\
      &G_{\theta}\ket{11}\hspace{8pt} = \ket{11} \\
      &G_{\theta}\ket{10}\hspace{8pt} = \cos{(2\theta)}\ket{10} + \sin{(2\theta)}\ket{01} \\
      &G_{\theta}\ket{01}\hspace{8pt} = \cos{(2\theta)}\ket{01} - \sin{(2\theta)}\ket{10}
    \end{aligned}
  \end{cases} 
\end{equation}
We find that $G(\frac{\pi}{8})H_{2}G(-\frac{\pi}{8}) = Z_1 - Z_2$, which is diagonal in the computational basis. Hence, we can claim that
\begin{equation}
    G(-\frac{\pi}{8})\ket{\text{computational basis state}} = \ket{\text{eigenstate of $H_{XY}$}} \,.
\end{equation}
In Fig.~\ref{ham_diag_fig}, we illustrate how a series of Givens rotations can be used to go to and from the eigenstate basis of the Hamiltonian for the time evolution operator of the 4-spin chain Gibbs state $\rho_4$. In this case, $n = 4$, and hence, $\frac{n(n-1)}{2} = \frac{4(4-1)}{2} = 6$ Givens rotations are used.

\subsection{Specific heat measurements}\label{spec_heat_sec}
\begin{figure*}[ht!]
    
    \centering
    
    \includegraphics[width=\textwidth]{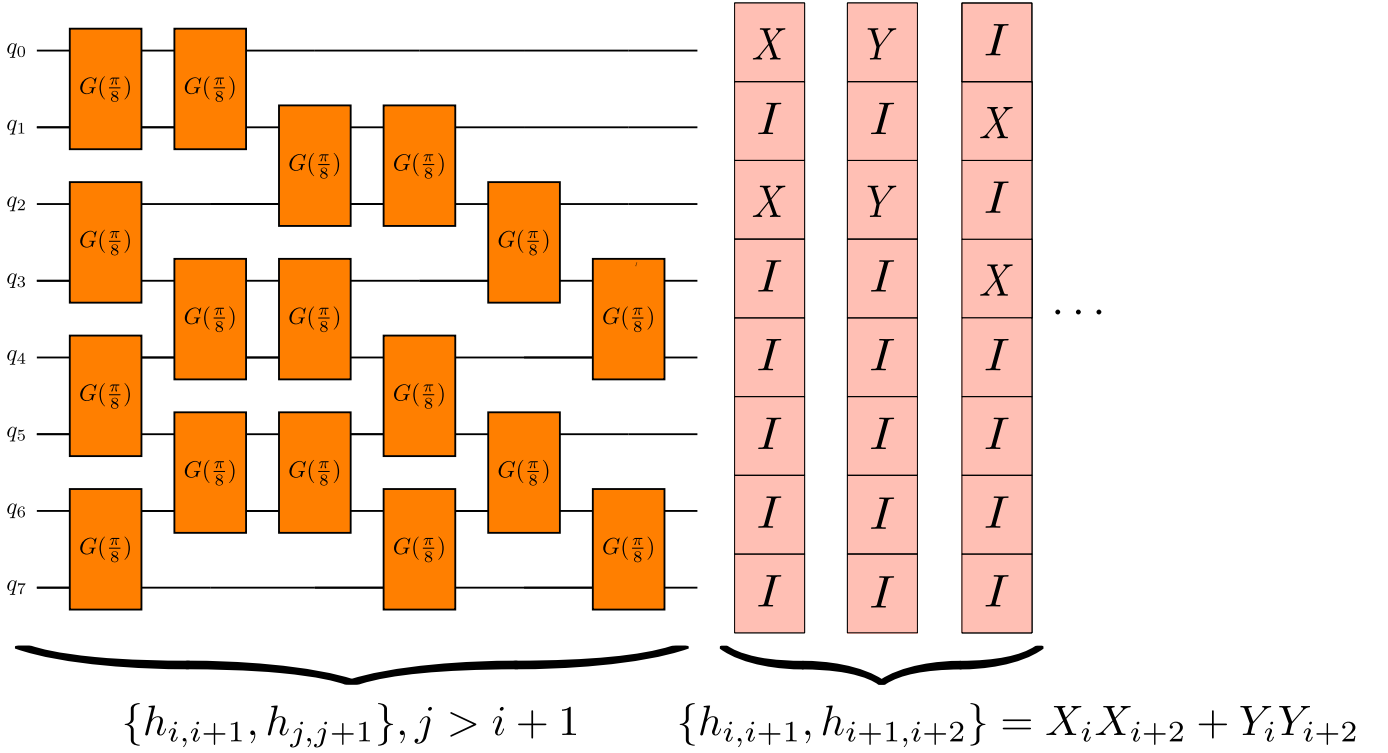}
    
    \caption{\textbf{The circuit configurations needed to measure $\boldsymbol{\langle H^2 \rangle}$ for $\boldsymbol{\rho_8}$:} Via specific configurations of the Givens rotations at $\theta = \frac{\pi}{8}$, we are able to prepare the anticommutator terms of the individual terms in the Hamiltonian $H$ with terms of the form $h_{i,i+1} = X_iX_{i+1} + Y_iY_{i+1}$. The results of such measurements are then weighted accordingly with respect to the initial state of the circuit, accounting for all of the terms needed for the measurement of the specific heat $C_\nu$.}
    \label{specheatcirc}
\end{figure*}

We aim to measure the specific heat $C_\nu$ of the global Gibbs state $\rho_8$ for both the 1D XY Hamiltonian model and the 1D Heisenberg Hamiltonian model. This necessitates that we measure $\langle H \rangle$ and $\langle H^2 \rangle$ as per Eq.~\eqref{specheq}. Our measurements are executed via the Estimator primitive in Qiskit Runtime, where we first prepare our individual cluster states making up the expansion, and then determine the objective observables to be measured, which would be the individual terms of $\langle H \rangle$ and $\langle H^2 \rangle$. For a Hamiltonian $H$ of the form 
\begin{equation}
    H = \sum^{n-1}_{i = 1} h_{i,i+1}\,,
\end{equation}
where $h_{i,i+1} = X_i X_{i+1} + Y_i Y_{i+1}$ in the case of the XY model and $n = 8$ in the case of $\rho_8$, we can prove that $H^2$ is written as
\begin{equation}
    H^2 = \sum^{7}_{i = 1} (h_{i,i+1})^2 + \sum^{6}_{i = 1} \{h_{i,i+1}, h_{i+1,i+2}\} + \sum^{7}_{\substack{i,j\\
    \abs{i-j}> 1}} \{h_{i,i+1}, h_{j,j+1}\} \,.
\end{equation} 
This calls for a huge number of circuits if we would implement a circuit for each individual observable, especially for the last term of $\langle H^2 \rangle$ where there is a large number of cross-terms we must account for. However, with the use of the Givens rotations, we are capable to rotating to the eigenstate basis to be able to capture more terms within specific configurations as illustrated in Fig.~\ref{specheatcirc}.

\subsection{IBM devices used for results}
We now present some of the calibration details of the IBM devices that were used to produce our results.

\begin{table}[htp!]
\centering
\begin{tabular}{||c | c | c | c | c||} 
 \hline
 Qubit $\#$: & $q_2$ & $q_1$ & $q_4$ & $q_7$ \\ [0.5ex] 
 \hline\hline
 Frequency (GHz) & 5.053 & 4.836 & 4.959 & 4.883\\ 
 $T_1 (\mu s)$ & 93.021 & 116.810 & 171.016 & 151.802 \\
 $T_2 (\mu s)$ & 248.060 & 96.011 & 196.874 & 166.908 \\
 Readout assignment error & 0.0585&0.0053& 0.0062 & 0.0087 \\ [1ex] 
 \hline
\end{tabular}
\caption{\textbf{Calibration details of \textit{ibm\_algiers}}: Listed in this table are some of the calibration details of \textit{ibm\_algiers} when using it to produce the correlation functions for the XY model Hamiltonian for the $\rho_4$ Gibbs state results in Figs.~\ref{rh04_im},\ref{rho4_panels}, and \ref{sqw_panels}. It is also to be noted that the CNOT error between the qubits was as follows: 2-1: 0.0055; 1-4:0.0055; 4-7: 0.0054.}
\label{table:2}
\end{table}

\begin{figure}[htp!]
    \centering
    \includegraphics[width=0.6\linewidth]{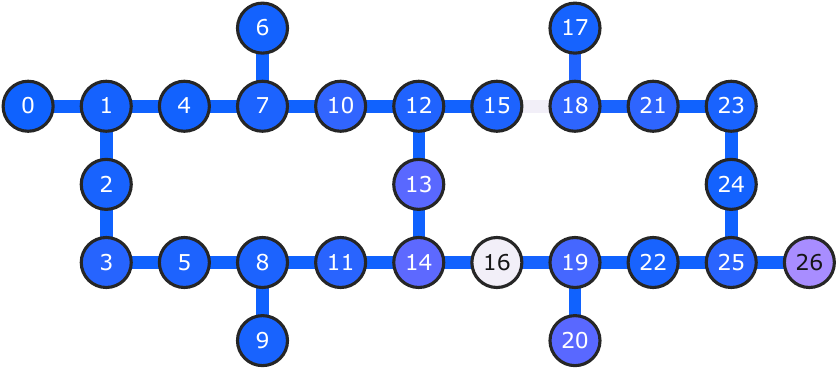}
    \caption[ibm\_algiers]{\textit{ibm\_algiers}: A 27-qubit IBM quantum device. }
    \label{cusco}
\end{figure}

\end{document}